\documentstyle[twocolumn,aps,prd,epsf,epsfig]{revtex}

\begin{document}
\twocolumn[\hsize\textwidth\columnwidth\hsize
\csname@twocolumnfalse%
\endcsname
\draft
\title{Waves on an interface between
two phase-separated Bose-Einstein condensates}
\author{I.E.Mazets}
\address{{\setlength{\baselineskip}{18pt}Ioffe Physico-Technical
Institute, 194021 St.Petersburg, Russia}} 
\maketitle

\begin{abstract}
We examine waves localized near a boundary between two weakly
segregated Bose-Einstein condensates. In the case of a wavelength
of order of or larger than the thickness of the overlap region the
variational method is used. The dispersion laws for the two
oscillation branches are found in analytic form. The opposite case
of a wavelength much shorter than the healing length in the bulk
condensate is also discussed.\\ 
\end{abstract}
\vskip1pc]

The studies of mixtures of two superfluid Bose liquids ascend to
the early work by Khalatnikov \cite{khal} who determined the
possible sound modes in such a system within a macroscopic
(hydrodynamical) approach. Later a microscopic theory was
developed: Bassichis \cite{qb} considered a neutral mixture of two
charged Bose gases with Coulomb interactions, and Nepomnyashchii
\cite{nep76} considered a system composed of bosons of two kinds
interacting via short-range potentials. In Ref.\cite{nep76}, most
closely related to the case of a two-component degenerate dilute
atomic vapor, the two-branch spectrum of elementary excitations
and following from it criterion for stability of the ground state
were found.

The advances in experiments on Bose--Einstein condensation of
trapped alkali atoms gave rise  to an  interest to this subject.
The ground state configuration of a two-component atomic Bose--Einstein
condensate (BEC) in the presence of a harmonic confining potential
at zero temperature
was calculated by Ho and Shenoy \cite{trap96} in the Thomas--Fermi limit.
The collective oscillation frequencies for trapped binary BECs were
also determined \cite{j98,f98,a98}. Interesting numerical results are
obtained by Pu and Bigelow \cite{pb}.
These studies reveal that if
the number of trapped atoms of each kind is large enough, the
ground state physics can be qualitatively understood from the
arguments valid for a case of absence of a trap \cite{nep76}.
In the latter case, the ground state properties are determined by a
certain relation between the coupling constants. Namely,
if the intercomponent repulsion (we do not consider attractive
potentials in the present paper)
is small enough, i.e., if
$g_{12}<\sqrt{g_{11}g_{22}}$, where
$g_{ij}=2 \pi \hbar [m_im_j/(m_i+m_j)]^{-1}a_{ij}$, $a_{ij}$ is the
corresponding {\it s}-wave scattering length of a pair of ultracold
atoms when one of them is of the $j$th kind and another is of the
$i$th kind, $m_j$ is the mass
of an atom for the {\it j}th component of the mixed BEC, $i\, j=1,\,2$,
then the two degenerate Bose gases are
miscible, i.e. they co-exist in all the volume.
In the opposite case, $g_{12}>\sqrt{g_{11}g_{22}}$, the two components
are immiscible and separated in space. In the latter case, a
new physics related to the intercomponent boundary arises. The
steady-state energetics of such an interface was estimated by
Timmermans \cite{if1}. Ao and Chui \cite{if2} performed more detailed
analysis, in particular, they found that there are two different
regimes called, correspondingly, weakly and strongly segregated phase.
The former one takes place if $g_{12}$ exceeds $\sqrt{g_{11}g_{22}}$
only slightly. In such a case the component interpenetration depth is
quite large and proportional to $\gamma ^{-1/2}$, where
\begin{equation}
\gamma =g_{12}(g_{11}g_{22})^{-1/2} - 1 .                 \label{ga}
\end{equation}
The latter limiting case can be achieved provided that $\gamma \gg 1$,
under such a condition the density profile at the intercomponent
boundary has quite a different shape and its thickness is determined by
individual healing lengths for the bulk condensates.

There are also experimental works on two-component BECs made by the JILA
and NIST group \cite{jilaj97,jilaa98} and by the MIT group \cite{mit98}.
The JILA and NIST group created binary mixtures of ultracold bosons by
populating different magnetic sublevels, $\left| F,\, m_F\right \rangle $
of $^{87}$Rb. It is worth to note that a mixture of the
$\left| 2,\, 2\right \rangle $ and $\left| 1,\, -1\right \rangle $
exhibits behaviour of a miscible system \cite{jilaj97}
(cf. the related theoretical paper \cite{bk}), and the mixture of the
$\left| 2,\, 1\right \rangle $ and $\left| 1,\, -1\right \rangle $
looks like an example of a weakly segregated system \cite{jilaa98}.

In the present paper, we consider normal modes for intercomponent
boundary oscillations in the weakly segregated phase regime, i.e., for
$0<\gamma \ll 1$, since a study of excited states of an immiscible system
was lacking up to now. The Bogoliubov spectrum of a translationally
invariant mixed BEC was obtained in Ref.\cite{nep76}, but it revealed
only the fact that such a homogeneous
system is dynamically unstable against
long-wavelength perturbations and, hence, tends to become phase-separated.
One can think that dynamics of a phase separated system with an
interface between the components 1 and 2 in the long-wavelength regime 
is adequately described by a
concept of a surface tension with the surface tension constant inferred
from the ground state properties \cite{if1}. However, a rigorous proof of 
this result was still lacking. In the present paper, we fill this gap and 
examine also the opposite (short-wavelength) regime. 

Let us introduce two macroscopic wave functions for the components
$j=1,2$ as $\sqrt{n_j} \Phi _j({\bf r},t)\exp (-i\mu _jt/\hbar )$,
where $\mu _j=\hbar g_{jj}n_j$
is the chemical potential of the corresponding component,
and $n_j$ is its bulk number density. In the
present paper, we consider a case of absence of any external potential,
so the coupled set of the Gross--Pitaevskii equations
\cite{j98,if1} is reduced to
\begin{eqnarray}
i\dot \Phi _1&=&-\frac \hbar {2m_1}\nabla ^2\Phi _1-\frac {\mu _1}\hbar
\Phi _1+g_{11}n_1\left| \Phi _1\right| ^2\Phi _1 +\nonumber
\\ & & g_{12} n_2  \left| \Phi _2\right| ^2\Phi _1,
\label{phit}\\
i\dot \Phi _2&=&-\frac \hbar {2m_2}\nabla ^2\Phi _2-\frac {\mu _2}\hbar
\Phi _2+g_{22}n_2\left| \Phi _2\right| ^2\Phi _2 +\nonumber
\\ &  & g_{12} n_1 \left| \Phi _1\right| ^2\Phi _2. \label{phyt}
\end{eqnarray}
We consider a case of two immiscible BECs.
We denote the ground state solution for $\Phi _j$ by $\phi _j({\bf r})$.
Let the first component occupy a semi-infinite space left to the $z=0$
plane that can be associated with the intercomponent boundary, i.e.,
$\phi _1\rightarrow 1$ if $z\rightarrow -\infty $
and $\phi _1\rightarrow 0$ if $z\rightarrow +\infty $. Respectively,
$\phi _2\rightarrow 0$ if $z\rightarrow -\infty $
and $\phi _2\rightarrow 1$ if $z\rightarrow +\infty $.
Hydrostatic arguments lead to the equality of the bulk component
pressures \cite{if1}:
\begin{equation}
\frac 12g_{11}n_1^2=\frac 12g_{22}n_2^2.            \label{hst}
\end{equation}

To get a qualitatively correct physical picture, it is sufficient to
consider a symmetric case: $m_1=m_2$ and $g_{11}=g_{22}$. Then we can
introduce the time unit $(g_{11}n_1)^{-1}$ and the length unit
$\sqrt{\hbar /(2m_1g_{11}n_1)}$ and write Eqs.(\ref{phit},~\ref{phyt})
in the dimensionless form (hereafter, for the sake of brevity, we do not
introduce  specific notation for the dimensionless time and
coordinate):
\begin{eqnarray}
i\dot \Phi _1&=&-\nabla ^2\Phi_1-\Phi _1+\left| \Phi _1\right| ^2\Phi _1+
(1+\gamma )\left| \Phi _2\right| ^2\Phi _1,
\label{phad}\\
i\dot \Phi _2&=&-\nabla ^2\Phi_2-\Phi _2+\left| \Phi _2\right| ^2\Phi _2+
(1+\gamma )\left| \Phi _1\right| ^2\Phi _2.
\label{phda}
\end{eqnarray}
Obviously, the steady state solution of Eqs.(\ref{phad},~\ref{phda})
possesses the
following symmetry:
\begin{equation}
\phi _1(z)=\phi _2(-z)   .                           \label{zmz}
\end{equation}

Now we have to determine the steady state order parameters for the
two components of the immiscible BEC. This task has been recently
solved numerically \cite{nu01}. However, we demonstrate that under
the above mentioned symmetry conditions it is possible to get a
reasonable approximation in the analytic form. First of all, let
us represent the order parameters as $$ \phi _1=\varrho \cos
\alpha , \quad \phi _2=\varrho \sin \alpha .$$  In the steady
state, the set of Eqs.(\ref{phad},~\ref{phda}) can be than
rewritten as
\begin{eqnarray}
\varrho ^{\prime \prime }&=&-\varrho +\varrho ^3+\frac \gamma
2\varrho ^3\sin ^2(2\alpha )+\varrho \alpha ^{\prime \,2},
\label{eqvr}  \\ \left( \varrho ^2\alpha ^\prime \right) ^\prime
&=&\frac \gamma 2\varrho ^4\sin (2\alpha )\cos (2\alpha ) .
\label{eqal}
\end{eqnarray}
Here prime denotes a derivative with respect to $z$. The boundary
conditions are
\begin{equation}
\alpha (-\infty)=0,\quad \alpha (+\infty )=\frac \pi 2, \quad \varrho
(\pm \infty )=1. \label{bcra}
\end{equation}
Since, by assumption, $\gamma $ is a small parameter, we use the
following decomposition in series: $$\varrho =\sum _{n=0}^\infty
\varrho ^{(n)}, $$ where $\varrho ^{(n+1)}/\varrho ^{(n)} =\cal
O(\gamma )$. The second unknown function, $\alpha $, can be
treated in a similar way. It is quite easy to find a solution in the
lowest order approximation, when we 
omit the third and fourth terms in the right hand side of Eq.(\ref{eqvr}) 
and substitute $\varrho ^{(0)} $ instead of $\varrho $ into 
Eq.(\ref{eqal}):
\begin{equation}
\varrho ^{(0)}=1, \quad \alpha ^{(0)}=\arctan \,
e^{\sqrt{\gamma}z}.       \label{zthord}
\end{equation}
The ''slowly varying co-ordinate'' $\sqrt{\gamma }z$ is treated
here as a quantity of order of unity. Then we find from
Eq.(\ref{eqvr}) the first order correction to $\varrho $. If we
substitute $\varrho = \varrho ^{(0)}+\varrho ^{(1)}$ into
Eq.(\ref{eqvr}) and linearize this equation with respect to
$\varrho ^{(1)}$ we get $$\varrho ^{(1)\, \prime \prime }-2\varrho
^{(1)}=3\gamma \cos ^2\alpha ^{(0)}\sin ^2\alpha ^{(0)} .$$ Then
it is important to note that $\varrho ^{(1)\, \prime \prime }$  is
of order of $\gamma ^2$ rather then of $\gamma ^1$ and, hence,
have to be omitted when we find the first order correction which,
finally, can be written as
\begin{equation}
\varrho ^{(1)} =-\frac {3\gamma }2\frac {e^{2\sqrt{\gamma
}z}}{\left( 1+e^{2\sqrt{\gamma }z}\right) ^2}       . \label{r1}
\end{equation}
Note, that only zeroth order approximation for $\alpha $ is
necessary for determination of $\varrho ^{(1)}$.

After Eq.(\ref{r1}) has been obtained, we can find the first order
correction to $\alpha $. We approximate the latter quantity by the
following formula:
\begin{equation}
\alpha =\arctan e^{\sqrt{\gamma }z+g(\sqrt{\gamma }z)},
\label{ASw1} \end{equation}
where $g=\cal O(\gamma )$.
Substituting this approximation into Eq.(\ref{eqal}) where
$\varrho ^{(0)}+\varrho ^{(1)}$ stands for $\varrho $, after some
tedious calculations we obtain
\begin{equation}
g =\frac {3\gamma }{16}\tanh (\sqrt{\gamma }z).
\label{ASw2}
\end{equation}

Thus the steady state order parameter for the 1st component of the
BEC with the $\cal{O}$ $(\gamma ^2)$ accuracy reads as
\begin{equation}
\phi _1(z)= \frac 1{\sqrt{1+e^{2\zeta }}} \left[ 1 -\frac {3\gamma
}2 \frac {e^{2\sqrt{\gamma }z}}{(e^{\sqrt{\gamma }z}+
e^{-\sqrt{\gamma }z})^2}\right] , \label{stst}
\end{equation}
where
\begin{equation}
\zeta =\sqrt{\gamma }z+\frac {3\gamma }{16} \tanh (\sqrt{\gamma
}z).     \label{ztzt}
\end{equation}
To find $\phi _2(z)$, one simply have to apply Eq.(\ref{zmz}). The 
analytic approximation given by Eqs.(\ref{stst},~\ref{ztzt}) is 
justified by numerical calculations. The difference between the 
0 analytic and numerical solutions behaves as 
$\cal O(\gamma ^2)$ in a wide range of $\gamma $ less than 0.1.

To investigate the long-wavelength oscillations, we apply the
variational method analogous to that used in Ref.\cite{smp} to study
the surface oscillations of a dense BEC in a trap. This method
implies minimization of the action functional associated with the
Lagrangian
\begin{eqnarray}
{\cal L}&=&\int d^3{\bf r}\, \left[ -\frac i2 (
\Phi _1^*\dot \Phi _1-\Phi _1\dot \Phi ^*_1+
\Phi _2^*\dot \Phi _2-\Phi _2\dot \Phi ^*_2)+\right. \nonumber \\ & &
(\nabla \Phi _1)(\nabla \Phi _1^*)+(\nabla \Phi _2)(\nabla \Phi _2^*)+
\frac 12 \Phi ^*_1\, ^2\Phi _1^2+     \label{lag}    \\   &  &
\left.  \frac 12 \Phi ^*_2\, ^2\Phi _2^2+
(1+\gamma )\Phi _1^*\Phi _2^*\Phi _1\Phi _2-\Phi _1^*\Phi _1 -
\Phi _2^*\Phi _2 \right]             . \nonumber
\end{eqnarray}
For long-wavelength perturbations, we represent the test functions
as 
\begin{eqnarray} 
\Phi _j&=&\left[ \phi _j(z)+\beta _j(t)\cos (kx)h_j(z)\frac d{dz}
\phi _j(z)\right] \times \nonumber \\  &  & \exp [i\xi
_j(t) f_j (z)\cos (kx)], \label{test}
\end{eqnarray}
$j=1,2$. 
The quantity $-\beta _j(t)h_j(z)\cos (kx)$ can be interpreted as a 
displacement in the $z$-direction of an element of a quantum gas that 
in the stationary case has the density $\phi _j^2(z)$. The gradient of 
the phase $\xi _j(t)f_j(z)\cos (kx)$ multiplied by two gives the 
velocity of the oscillating quantum gas.  

Due to the problem symmetry,
\begin{equation} h_1(z)=h_2(-z). \label{h12}  \end{equation} 
The continuity equation implies that
\begin{equation}
h_j(z)\frac d{dz}\phi _j^2(z)=-2\frac d{dz}\left[ \phi _j^2(z)
\frac d{dz}f_j(z)\right] +2k^2\phi _j^2(z)f_j(z), \label{fth}
\end{equation}
$j=1,\, 2$. 
It is easy to see that it follows from Eqs.(\ref{h12},~\ref{fth}) that 
\begin{equation}f_1(z)=-f_2(-z).    \label{f12}    \end{equation}  

In the long-wavelength case, when 
\begin{equation} 
k\ll \sqrt{\gamma },         \label{lowa}
\end{equation}
the function $h_j(z)$ varies on a spatial scale of order of $k^{-1}$. 
From the other hand, since $h_j(z)$ appears in Eq.(\ref{test}) only 
multiplied by $d\phi _j^2(z)/dz$, one needs to know the exact 
behaviour of $h_j(z)$ only for 
$\left| z\right| \, ^<_\sim \, \gamma ^{-1/2}$, 
provided that $h_j(z)$ 
is finite outside this  small range of $z$. It means that, 
because of Eq.(\ref{lowa}), one can choose 
\begin{equation} 
h_j(z)=1.                         \label{edin}
\end{equation}

To find the quantity $f_j(z)$, we use the technique of tailoring 
asymptotics. Namely, for large negative $z$ (let say, for 
$z\, ^<_\sim \, -\gamma ^{-1/2}$), where $\phi _1^2(z)\approx 1$, 
Eq.(\ref{fth}) is reduced to 
$$
\frac {d^2}{dz^2}f_1(z)=k^2f_1(z), 
$$
and its solution approaching zero when $z\rightarrow -\infty $ is 
$f_1(z)=C_1e^{kz}$. In the opposite case, in the range of $z$ 
where $\phi _1(z)$ rapidly decreases, the 
second term in the right hand side of Eq(\ref{fth}) is negligible 
compared to the first one, because of the condition Eq.(\ref{lowa}), 
and, hence, $f_1(z)= C_2-z/2$. The matching condition for these 
two asymptotics and their first derivatives allows us to determine the 
constants $C_1$ and $C_2$. Thus, in the subsequent calculations 
we use the following expression 
\begin{equation}
f_1(z)=
-\frac 1{2k}e^{kz}\Theta (-z) -\left( \frac 1{2k}+\frac z2\right) 
\Theta (z)   ,  
\label{ffsf}
\end{equation}
where $\Theta (z)$ is the Heaviside's step function. This expession and its 
first derivative with respect to $z$ are continuous everywhere, including 
the point $z=0$.  
The result of Eq.(\ref{ffsf}) is also justified by numerical calculations; 
$f_2(z)$ is given by Eq.(\ref{f12}).  

Then we write the Lagrange equations for the generalized coordinates
$\beta _j(t)$, $\xi _j(t)$. First,  two of these equations enable
us to exclude the two variables $\xi _j=\dot \beta _j$. After this
has been done, we arrive at the two  equations containing $\beta _j$,
$\ddot \beta _j$ and describing two coupled harmonic oscillators.
The normal modes correspond to the two cases,
$$
\beta _1=\varsigma \beta _2, \qquad \varsigma =\pm 1.
$$
For $k\ll \sqrt{\gamma }$ their eigenfrequencies 
$\omega _\varsigma $  are given by 
\begin{eqnarray}
\omega _\varsigma ^2&=&2
\int _{-\infty }^\infty dz \,\Big {\{ } 
2(1+\gamma )(1-\varsigma )P(z)P(-z) +    \nonumber  \\ & & 
k^2 \left[ \frac d{dz}
\phi _1(z)\right] ^2 
\Big {\} }\bigg / 
\left[ \int _{-\infty }^\infty dz\, f_1(z)
\frac d{dz}\phi _1^2(z)\right] 
,         \label{reslwl}
\end{eqnarray}
where $P(z)=d\phi _1^2(z)/dz$. 

If the exact ground state wave function is used in Eq.(\ref{reslwl}),
then $\omega _\varsigma $
is a rigorous upper bound to the oscillation frequency for the
normal modes with no nodes along the {\it z}-axis. Expanding the 
right hand side of Eq.(\ref{reslwl}) in
series in $\gamma $ as well as in $k\gamma ^{-1/2}$ and 
retaining the leading terms, we obtain the following 
expressions for the lowest oscillation frequencies for two 
weakly segregated BECs in the long wavelength regime: 
\begin{eqnarray}
\omega _{+1}^2&=&\sqrt{\gamma }\,k^3, \label{opl} \\
\omega _{-1}^2&=&\frac 43 \sqrt{\gamma }\,k.  \label{omn}
\end{eqnarray}  

Now we can clarify the physical meaning of the two branches of the
dispersion law Eq.(\ref{reslwl}). The $\varsigma =+1$ branch
corresponds to in-phase oscillations of the two components and
represents interface bending. The wave number dependence, 
$\omega _{+1}\propto k^{3/2}$ for $k\rightarrow 0$, is specific for
this type of surface waves (capillary waves) \cite{hydr}. Also
these interface oscillations are an analog of soft modes \cite{k2}
in a mixture of miscible BECs with $\gamma $ being very small by
the absolute value. We call these oscillations ``in-phase''
because for them the $z$-projections of the velocities of both the
two components have the same sign. It is necessary to note that 
the value of $k^{-3}\omega _{+1}^2$ that is, in fact, the ratio of 
the surface tension constant to the sum of the bulk densities of the 
1st and 2nd BEC components is equal to $\sqrt{\gamma }$, i.e., 
displays the correct dependence on $\gamma $ following from the 
surface tension energy estimation given by Timmermans \cite{if1}.

The $\varsigma =-1$ branch
corresponds to the out-of-phase oscillations and, hence, to a time 
dependence of the depth of  
interpenetration of the two components. The
dispersion law, $\omega ^{(0)\,2}_{+1}\propto k$, is analogous to
that of surface waves on a boundary between a single component BEC
and vacuum \cite{smp}.

Now it is necessary to explain why in the present problem one cannot 
choose
\begin{equation}
f_1(z)=e^{kz},\qquad h_1(z)=-2ke^{kz},
\label{fh}
\end{equation}
as it is done in Ref.\cite{smp}. 
The reason is that if one applies Eq.(\ref{fh}) then for 
$z\rightarrow +\infty $ the difference between the steady-state 
order parameter $\phi _1(z)$ and the time-dependent order 
parameter $\Phi _1(x,z,t)$ is much larger (by the exponential factor 
$e^{kz}$) than $\phi _1(z)$. So, the condition of the relative 
smallness of the order parameter perturbation breaks down, and the 
validity of the approach described above cannot be guaranteed. 
In Ref.\cite{smp}, the condensate is assumed to fill only half of 
the space, and Eqs.(\ref{fh}) are used for $z<0$ only, while 
the condensate density for $z>0$ is identically zero, and no 
such a difficulty occurs. 

In our case, the wrong choice of $h_j(z)$ and $f_j(z)$ given by 
Eq.(\ref{fh}) results in the correct value for $\omega _{-1}^2$ for 
$k\ll \sqrt{\gamma }$ but substantially, 
by a factor of order of $\gamma ^{-1}$, overestimates the 
eigenfrequency for the in-phase mode. The latter result is apparently 
inconsistent with the surface tension estimations given in 
Ref.\cite{if1}. Moreover, use of Eq.(\ref{fh}), instead of 
Eqs.(\ref{edin},~\ref{ffsf}), leads to an 
erroneous conclusion about non-monotonous 
behaviour of the dispersion laws for both the in-phase and 
out-of-phase oscillation when $k$ approaches $\sqrt{\gamma }$. 
Certainly, the latter is simply an artefact of ineligible choice of 
the test function.

Another important limiting case is the case of very short wavelengths,
$k\gg 1$. Under this condition an elementary excitation is
very close to a single particle excitation (in other words, the $v$
coefficient of the Bogoliubov's transformation \cite{nnb} is small
compared to the $u$ coefficient). Thus, the oscillation frequency is
the eigenvalue of the Hamiltonian problem
\begin{eqnarray}
-&u_1^{\prime \prime }(z)+\left[2\phi _1^2(z)+(1+\gamma )
\phi _1^2(-z)-1+\right. 
&  \nonumber \\  &  \left. k^2\right]  u_1(z)+
(1+\gamma )\phi _1(z)\phi _1(-z)u_2(z) &=\omega u_1(z) ,
\label{hmt1}  \end{eqnarray}
\begin{eqnarray}
-&u_2^{\prime \prime }(z)+\left[   
2\phi _1^2(-z)+(1+\gamma )\phi _1^2(z)-1+\right. 
&   \nonumber \\  &  \left. k^2\right]  u_2(z)+
(1+\gamma )\phi _1(z)\phi _1(-z)u_1(z) &=\omega u_2(z) ,
\label{hmt2} 
\end{eqnarray}
It is obvious that there are two types of solutions of this 
set of coupled Scr\"odinger equations corresponding to the 
in-phase ($\varsigma =+1$) and out-of-phase ($\varsigma =-1$) 
oscillations: 
\begin{equation}
u_1(z)=-\varsigma u_2(-z).   \label{us} 
\end{equation}

The set of Eqs.(\ref{hmt1},~\ref{hmt2}) was solved taking into account 
the symmetry condition Eq.(\ref{us}) by the Hamiltonian diagonalization 
in the basis of harmonic functions $L^{-1/2}\sin [\pi m(z+L)/(2L)]$, 
where $m$ is a positive integer
number, and $L$ is chosen to be much larger than $\gamma ^{-1/2}$. 
The spectrum of the in-phase oscillations consists of the single 
discrete value equal to $k^2+0.32\gamma $ and the two branches 
of the continuous spectrum with the dispersion laws $k^2+1+Q^2$, $Q>0$,   
and $k^2+\gamma +q^2$, $q>0$. One branch corresponds 
to the incidence to the intercomponent boundary of 
sound waves from both the left and right sides, and $Q$ is the 
absolute value of the $z$-component of the wave vector of the 
incident wave. No atoms of one component are injected into the 
bulk of the condensate composed of atoms of another kind. 
In contrast, the second branch corresponds to injection of atoms of 
the 1st kind into the bulk of the condensate of atoms of the 2nd kind, 
and {\it vice versa}; $q$ is the $z$-projection of the momentum of 
an injected atom. 

\begin{figure}[h]
\begin{center}
\centerline{\psfig{figure=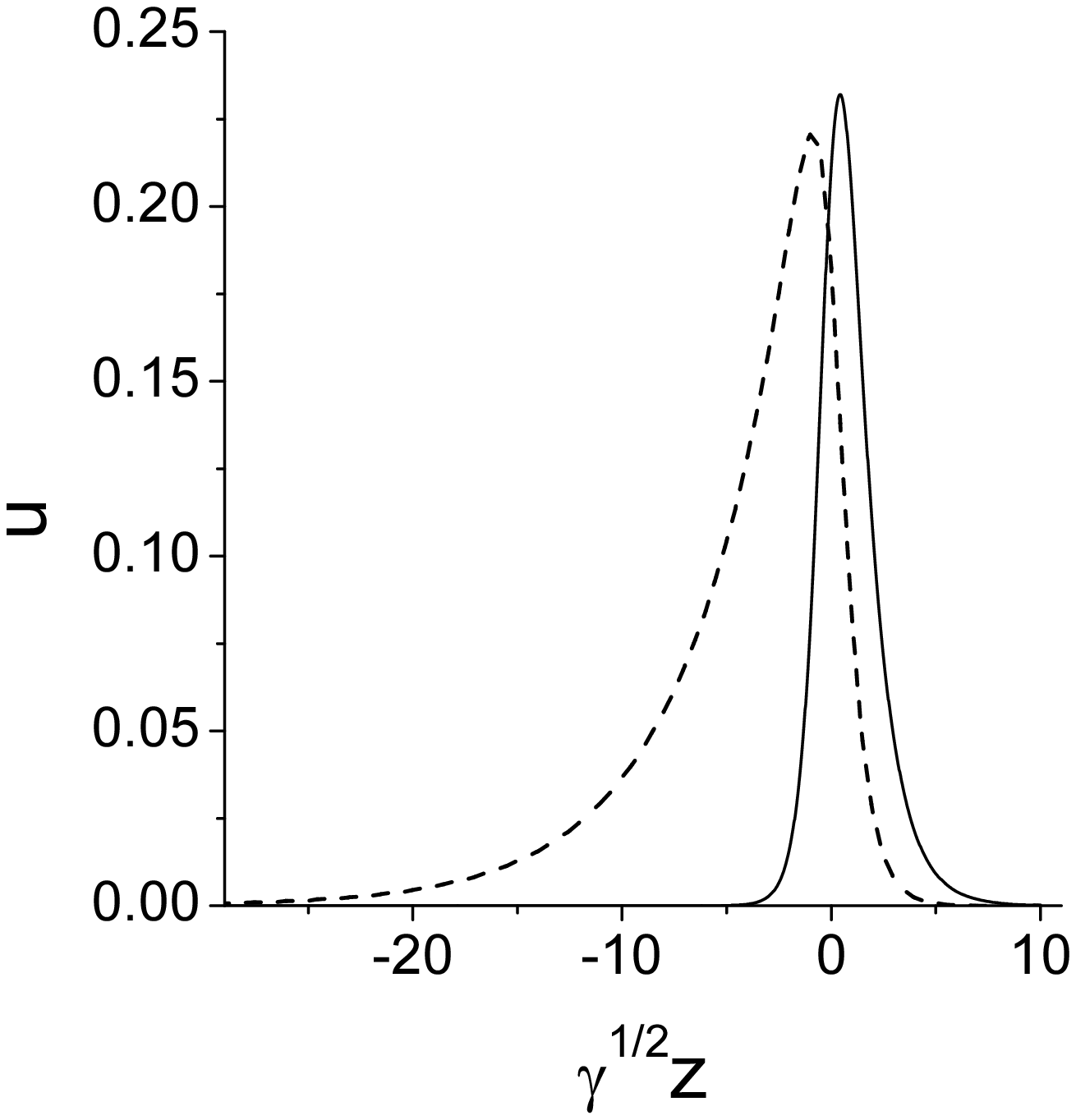,height=6cm}}
\end{center}
\vspace{-0.4cm}
\begin{caption}
{Numerical solutions of 
Eqs.(\ref{hmt1},~\ref{hmt2}) the in-phase (solid line) and
out-of-phase (dashed line) oscillations; $\gamma =0.01$; 
$z$-coordinate is dimensionless, the eigenfunctions are plotted in
arbitrary units. }
\end{caption}
\label{fkla}
\end{figure}

The spectrum of the out-of-phase oscillations contains again single 
discrete value equal to $k^2+1-0.044\gamma $ 
and the two branches of continuous spectrum. 
The dispersion laws and physical meanings of these two branches are 
the same as for similar branches in the case of in-phase oscillations. 
The fact that the discrete eigenvalue corresponding to the 
nodeless eigenfunction shown by the dotted line in Fig.1 
is not the minimum eigenvalue 
in this case looks surprising, nevertheless, it is quite natural. 
Indeed, if we substitute Eq.(\ref{us}) into Eq.(\ref{hmt1}) 
we get a 
problem with the Hamiltonian containing the inversion operator. 
And for a Hamiltonian of such a kind the usual theorem, saying that a 
nodeless eigenfunction corresponds to the lowest eigenvalue, does not 
apply, as can be seen from various exactly solvable examples. 

The eigenfunctions $u_\varsigma (z)$ corresponding to the discrete 
eigenfrequencies are shown in Fig.1. Note, that the out-of-phase 
oscillation is much localized than the in-phase one.

To summarize, we performed an analysis of dispersion of  waves on
a boundary between two weakly segregated dilute atomic BEC. For very
small wavenumbers $k$ we recover usual capillary and surface waves, with
frequency squared proportional to $k^3$ and $k$, respectively.
For large wavenumbers, each of the two types of oscillations  
(the in-phase and out-of-phase modes) has only one particular mode 
well localized in $z$-direction  near the boundary. Other surface 
modes are associated to simultaneous emission of either sound into 
the bulk condensate or to free particles of one kind to the BEC composed 
of atoms of another kind. The latter fact must make (due to the effect of 
quantum depletion) the depth of overlapping of the 
two weakly segregated ultracold bosonic clouds even larger than the 
simple mean-field theory predicts. This may be a very important 
cause of difficulties in an experimental determination of the 
parameter $\gamma $ \cite{jilaa98}.

The author is grateful to Prof. G.V. Shlyapnikov for fruitful
discussions. This work is supported by the Nederlandse Organisatie
voor Wetenschappelijk Onderzoek, project NWO-047-009.010, the
state program ``Universities of Russia'' (grant 015.01.01.04), and
the Ministry of Education of Russia (grant E00--3--12).

\end{document}